# Deep Learning Realm for Geophysics: Seismic Acquisition, Processing, Interpretation, and Inversion

Ping Lu

Anadarko Petroleum Corporation, Houston, TX, 77380

## Abstract

Applying deep-learning models to geophysical applications has attracted special attentions during the past a couple of years. There are several papers published in this domain involving with different topics primarily focusing on synthetic data. Based on the presented results, reaching the satisfaction from geophysicists is largely far away the performance of the conventional algorithms. Moreover, it is extremely hard to find documents with detailed illustrations about what are the best practices that researchers should follow regarding how to design appropriate deep-learning models to effectively and precisely tackle problems relevant to the field data. This paper serves as a summarization to demonstrate successful stories and share with extensive experiences we have gained during past several years in the process of designing and deploying of deep-learning models to the geophysical projects in a large scale. Four different disciplines are discussed individually with seismic acquisition, processing, interpretation, and inversion. Finally, special attentions about designing an effective deep-neural-networks especially for geophysics are discussed.

## Introduction

With the progressive development of artificial intelligence, geophysical community starts to realize the importance of leveraging the emerging technology to assist or even replace conventionally geophysical workflows. Three-dimensional seismic data, which is considered as the most important resource in assisting on the evaluation of the quality of reservoir and investment decision-making, has been heavily relied on and analyzed by geophysicists and geologists for a decade.

Given the fact that dozens of large-scale and complex attributes could be easily derived from the raw seismic data, human couldn't afford and don't have the ability to visualize and inspect such huge amount of information simultaneously. We ought to seek other opportunities in a way that deep-learning model emerges with the ability of providing the chance to quickly and precisely detect any targets from the seismic data. More importantly, there are still geophysical limitations associated with seismic signals. Given the factors of constrained budget and technologies, seismic imaging quality may not always reflect the true characterization of subsurface correctly and sharply, which could be caused by expensively computational cost and deficiency of mathematical algorithms. To reach the expectations from geophysicists, we still need to improve the technique from computational and theoretical point of view. Deep-learning model, which is able to mimic physics and may have the ability to go beyond physics, provides a totally new angle to resolve these challenges, which have been existing for decades and can't be solved from conventional approaches properly. Four different pillars support the entire deep-learning realm for assisting geophysics: acquisition, processing, interpretation, and inversion. Successful cases will be discussed individually in the following sections.

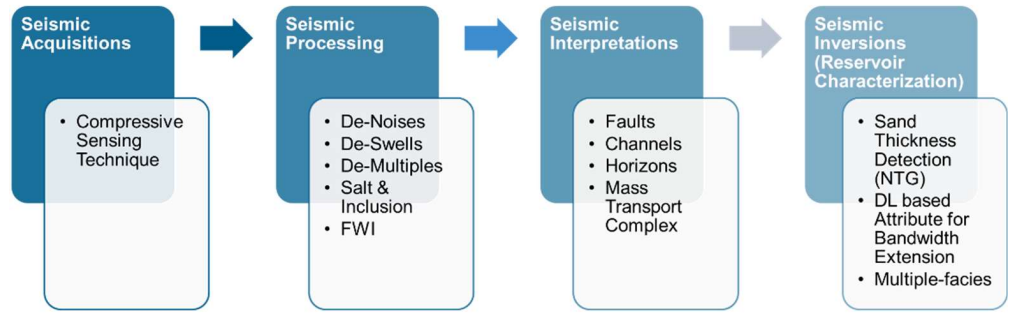

A generative model is a powerful way of learning any kind of data distribution using unsupervised learning, and it has achieved tremendous success in recent few years. All types of generative models aim at learning the true data distribution of the training dataset in order to generate new data points with some variations. Two of the most commonly used and efficient approaches are Variational Autoencoders (VAEs) and Generative Adversarial Networks (GANs). VAEs aim at maximizing the lower bound of the data log-likelihood, and GANs aim at achieving an equilibrium between Generator and Discriminator. GANs has been heavily relied on solving the tasks in overcoming geophysical limitations from our work. In addition, GANs provide a scalable framework for quantifying uncertainties in the analysis of reservoir characterizations due to the randomness or noise as the inputs.

A powerful classifier is desired for classifying label for each pixel. The evaluation metrics require to have extremely high precision in pixel-level, since missing one pixel from the inference may indicate of missing several meters for the subsurface characteristics. There are two different levels of interpretations needed for the exploration or production purposes. A structure interpretation that requires a lower precision versus a reservoir-level interpretation that requires the quantitative interpretation with much higher demanding on precision are carefully considered. Special attentions about designing an effective deep-neural-networks especially for geophysics are demonstrated at the end of the manuscript.

## Seismic Acquisition

**Compressive Sensing:**

Seismic imaging often suffers from aliasing artifacts if it is highly under-sampled. Conventional recovery relying on regularized iterative reconstructions, based on pre-defined sparsity transforms, usually involves time-consuming iterative optimizations, and may result into undesired artifacts such as over-smoothing. The use of deep-learning models as priors for compressive sensing tasks presents new potential for inexpensive seismic data acquisition.

There exist evidences that GANs excel in retrieving image content with improved quality and finer details, when comparing to conventional approaches. Therefore, a new framework, based on a GANs architecture trained on several historical surveys, is proposed to reconstruct missing seismic traces. The agility of the proposed approach lies in its ability to perfectly resolve the optimization limitation of conventional algorithms that are being used for inversion problems. Deep GANs, equipped with an appropriate loss function that essentially leverages the distribution of the entire survey, can serve as an alternative approach for tackling compressive-sensing problems with high precision and in a computationally efficient manner.

Advantages from using such an approach can be justified both by theoretical and experimental results. The robustness of the proposed approach on field data is validated by establishing the extent to which amplitudes and phase variations in the original data are faithfully preserved, while subsurface consistency is also achieved (Lu et al. 2019). Several applications to acquisition and processing, such as decreasing bin size, increasing offset and azimuth sampling, or increasing the fold, can directly and immediately benefit from adopting the proposed technique. While maintaining the same order of signal-to-noise ratio, GANs-based interpolation creates sharp, plausible images as shown in Figure 1.

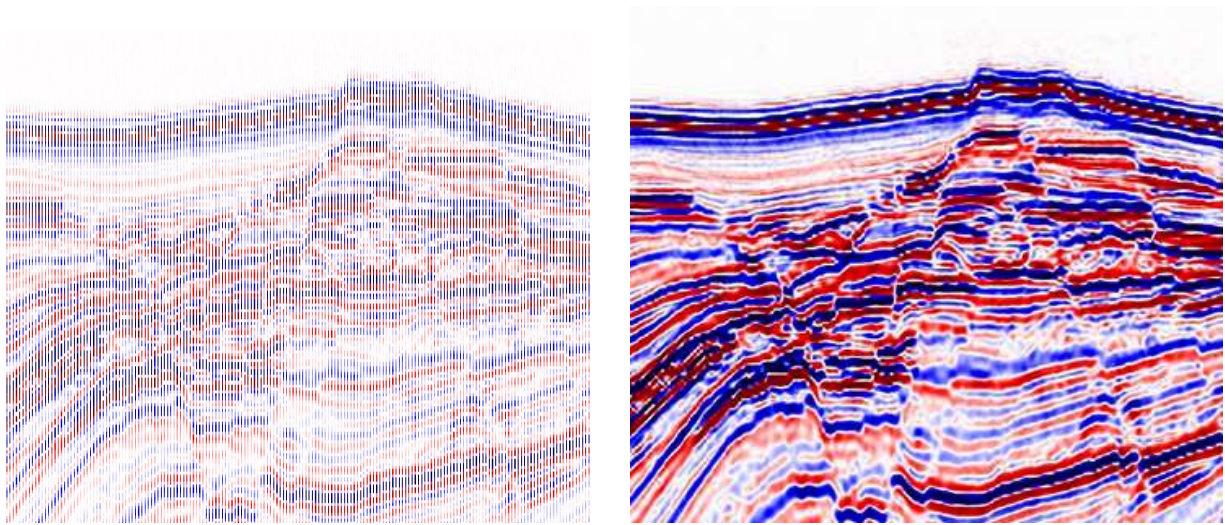

(a) Original seismic imaging　　　　　　　　　　(b) Imaging with filled traces

Figure 1. Comparisons of imaging qualities

Figure 2 presents a real case from Gulf of Mexico that interpolation based on generative adversarial networks produces better-sampled data sets, with stronger regularization and attenuated aliasing phenomenon, while providing greater fidelity on steep-dip events and amplitude-variation-with-offset analysis for migration.

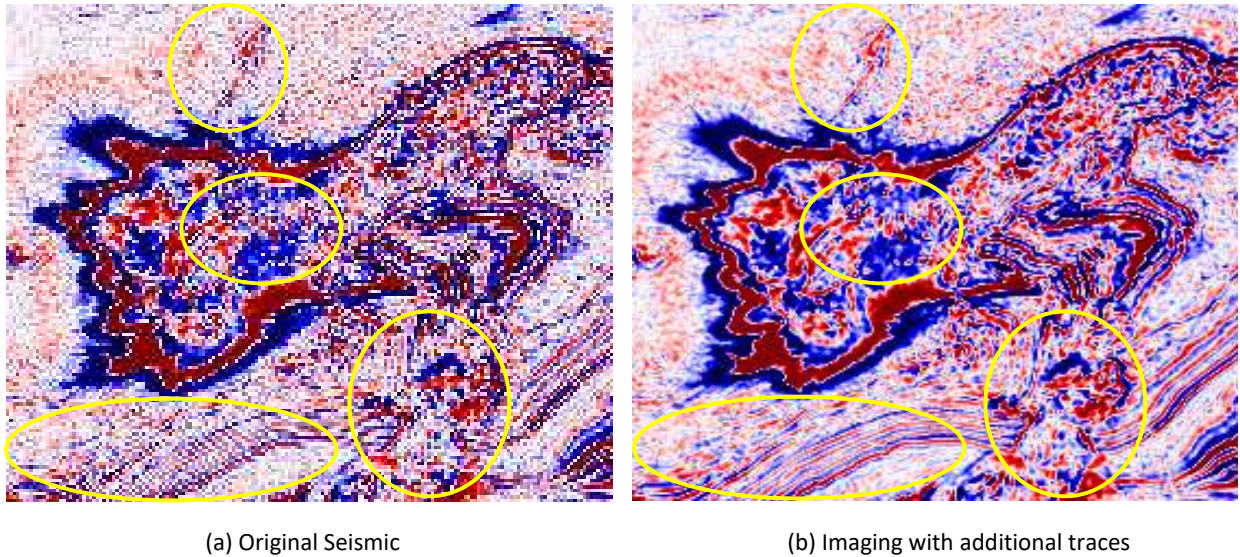

(a) Original Seismic  (b) Imaging with additional traces

Figure 2. Improvements of imaging qualities in depth

The primary goal of the proposed deep-learning model is to provide the foundations of an optimal design for seismic acquisition, without the loss in imaging quality. The development of optimization of survey layout largely relies on the progress of development of large-scale reconstruction algorithm for seismic wavelets. The performance of a reconstruction algorithm is judged on the basis of the level of compressive ratio, enhancement of imaging quality, attenuation of aliasing phenomenon, consistency between neighboring traces, preservation of original information, contamination of frequency spectrum, etc. The proposed GANs-based technique satisfies all these requirements and can be easily applied to optimize the nodes layout of seismic survey.

Two recommendations for non-uniform seismic survey grid, based on the evaluation of reconstructed seismic images and metrics, is proposed (Li et al. 2019). Along these lines, a compressive sensing design of a non-uniform grid over an asset in Gulf of Mexico, versus a traditional seismic survey grid which collects data uniformly at every few feet, is suggested, leveraging the proposed method. As the depth view shown in Figure 3, the recommended sampling points are densely distributed in regions with lithologic features and sparsely distributed in channelized regions. It successfully captures the heterogeneity of the seismic image. A well-optimized survey layout is eventually obtained by the proposed method which significantly contributes to seismic acquisition by reducing the number of necessary nodes.

| Ordinary Survey Grid | Optimized Survey Grid | Input | Reconstruction |
|---|---|---|---|
| Rcommendation 1 ||||
| 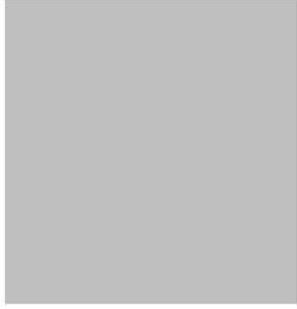 | 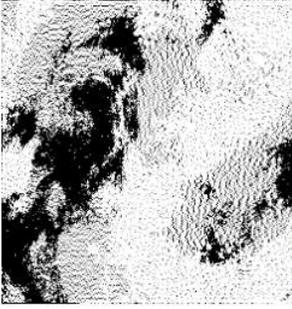 | 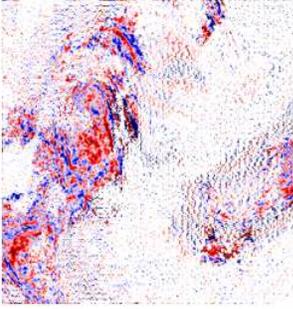 | 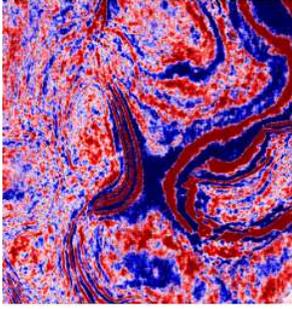 |
| Rcommendation 2 ||||
| 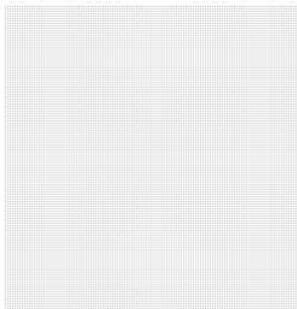 | 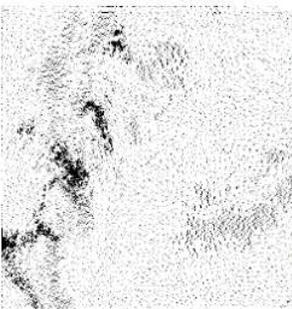 | 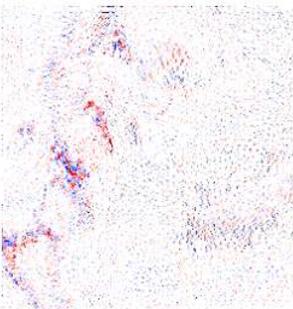 | 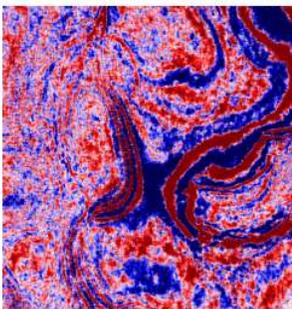 |

Figure 3. Depth view of recommended sampling 3D reconstruction on a region of the GoM dataset. Most of the information has been well reconstructed.

## Seismic Processing

### Noise Attenuation

In the geophysical field, seismic noise attenuation has been considered as a critical and long-standing problem, especially for the pre-stack data processing. We have proposed a totally new idea of leveraging the deep-learning model for this task. Rather than directly applying existing de-noising model from ordinary image to the seismic data, a particularly designed deep-learning model, based on residual neural networks, is introduced, with name as N2N-Seismic, which has a strong ability to recover the seismic signals back to intact condition with the preservation of primary signals. The proposed model, achieving with a great success in attenuating noise, has been tested on several seismic datasets. Certain metrics show that the presented method outperforms conventional approaches in terms of Signal-to-Noise (SNR) Ratio, Mean-Squared-Error (MSE), Phase Spectrum, etc. Moreover, robust tests in terms of effectively removing random noise from any dataset with strong and weak noises have been extensively scrutinized in making sure that the proposed model is able to maintain a good level of adaptation while dealing with large variations of noise characteristics and intensities (Zhao et al. 2019).

Having rigorous comparisons with conventional methods for several benchmark studies, our proposed deep-learning models successfully implement the tasks with a great success and achieve the prominent improvements in some respects. Figure 4 and 5 exhibit the de-noised results by the reference, f-x empirical mode decomposition, and proposed methods. It is apparent that, with the growth of the noise level, the de-noised image by the reference method loses more details, especially at the bottom edge. The main contribution of our work is to introduce a deep-learning solution for random noise attenuation tasks, which strongly outperforms conventional approach in several aspects. It is noticeable that such method absorbs benefits from the deep neural-networks in the computer vision applied to ordinary image denoising process, and also meets the geophysical requirements and expectations.

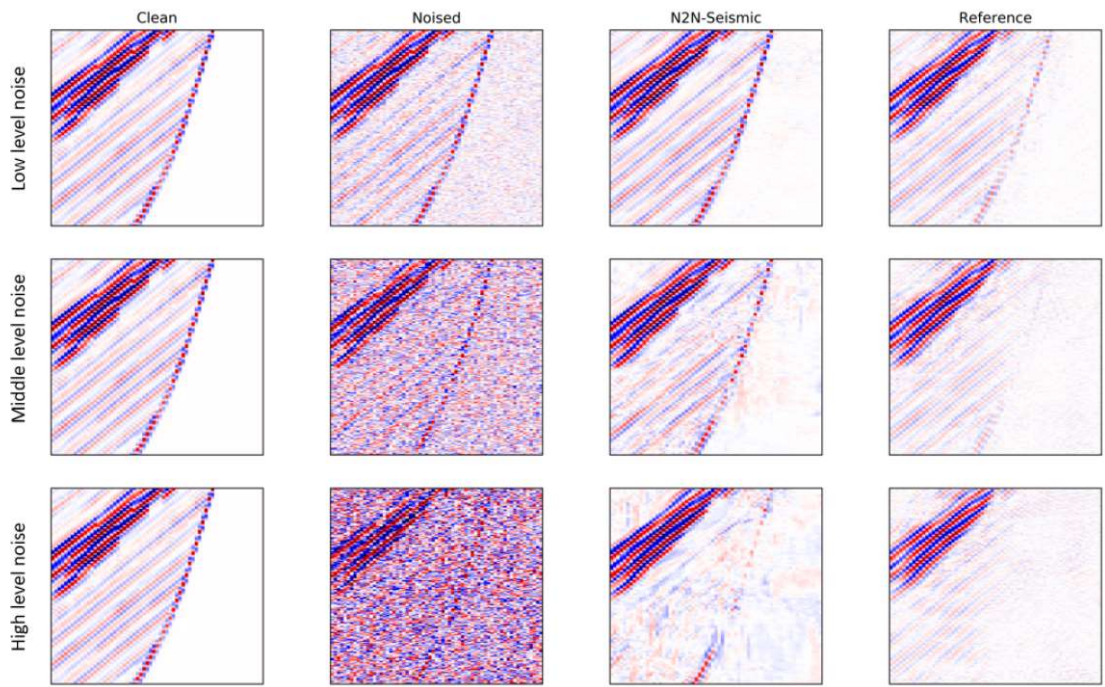

Figure 4. Performance comparison for random noise attenuation. Figure column (1) shows the partial clean image without random noise. Figure column (2) shows the partial clean image with low, middle, and high-level random noise. Figure column (3) is the de-noised seismic data by our N2N-Seismic methods, and Figure column (4) is the de-noised seismic data by the reference model.

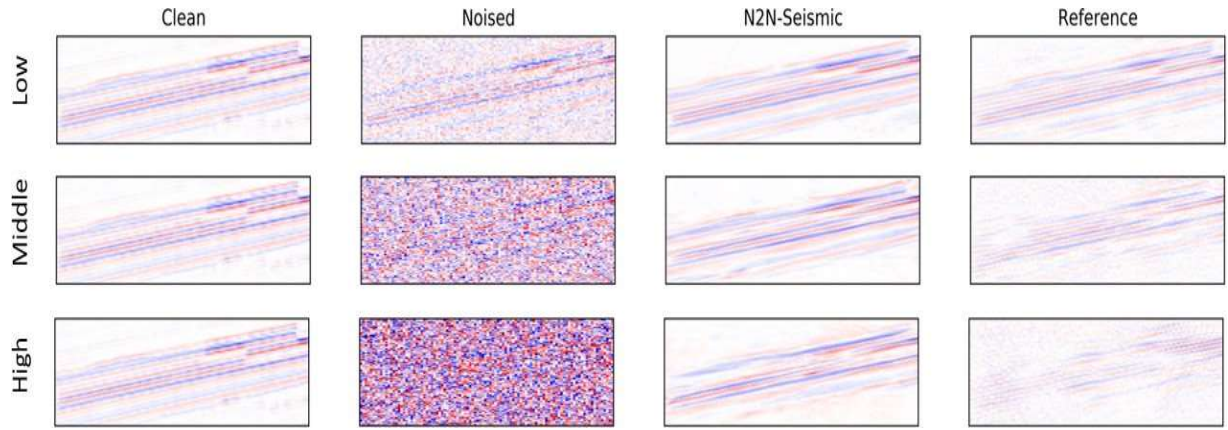

Figure 5. Performance comparison for random noise attenuation. Figure column (1) shows the partial clean image without random noise. Figure column (2) shows the partial clean image with low, middle, and high-level random noise. Figure column (3) is the de-noised seismic data by our N2N-Seismic methods, and Figure column (4) is the de-noised seismic data by the reference model.

## **Swell Attenuation**

In the geophysical domain, the attenuation of swell noise, caused by the long-period changes in the ocean surface, has been considered as another extremely challenging task, especially for the pre-stack seismic processing. The evaluation criteria for performance of an algorithm from the geophysical domain requires a high standard not only to obtain a high quality of visualized image, but also an ability to maintain the original seismic signals from the contaminated wavelets. In order to reach the eventual goal, we propose a totally new deep-learning approach, which is designed to effectively attenuate swell noises with different intensities and characteristics from gathers. Our model, Only2Noise, is applied on both a synthetic dataset and a field dataset, and obtains impressive results comparing with the conventional approach. Elimination of distortion of seismic wavelets during the separation process between swell noise and primary signals is totally guaranteed with the deep-learning model, which is trained with how to remove all contaminated swell noise from each frequency. With clear uplifts from two metrics of MSE and SNR, Only2Noise effectively and significantly enhances the quality of recovered data (Zhao et al. 2019).

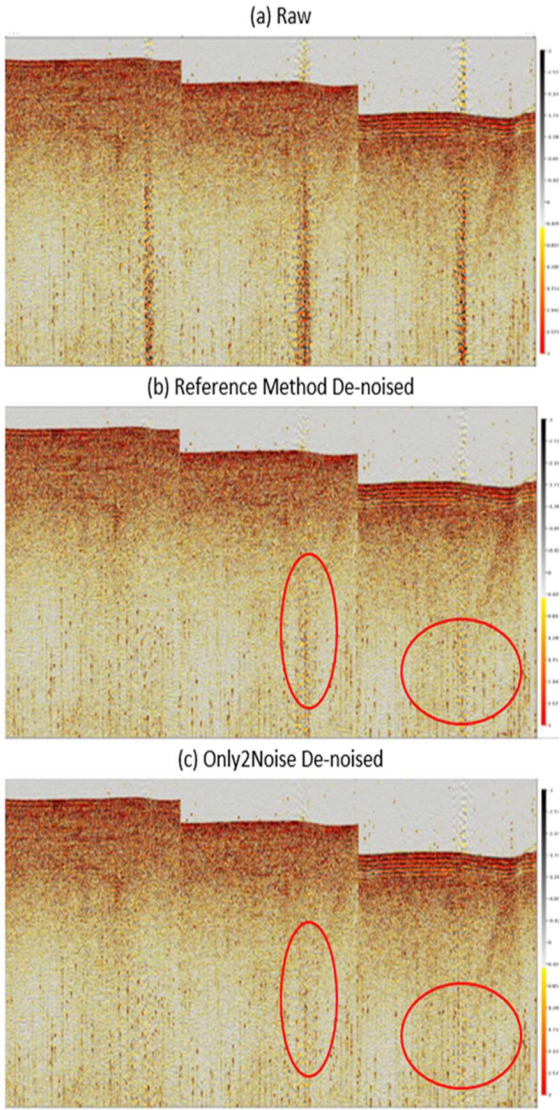

Figure 6. Field Data: Common channel comparison of corrupted raw data (a), Reference de-noised data (b), and Only2Noise de-noised data (c)

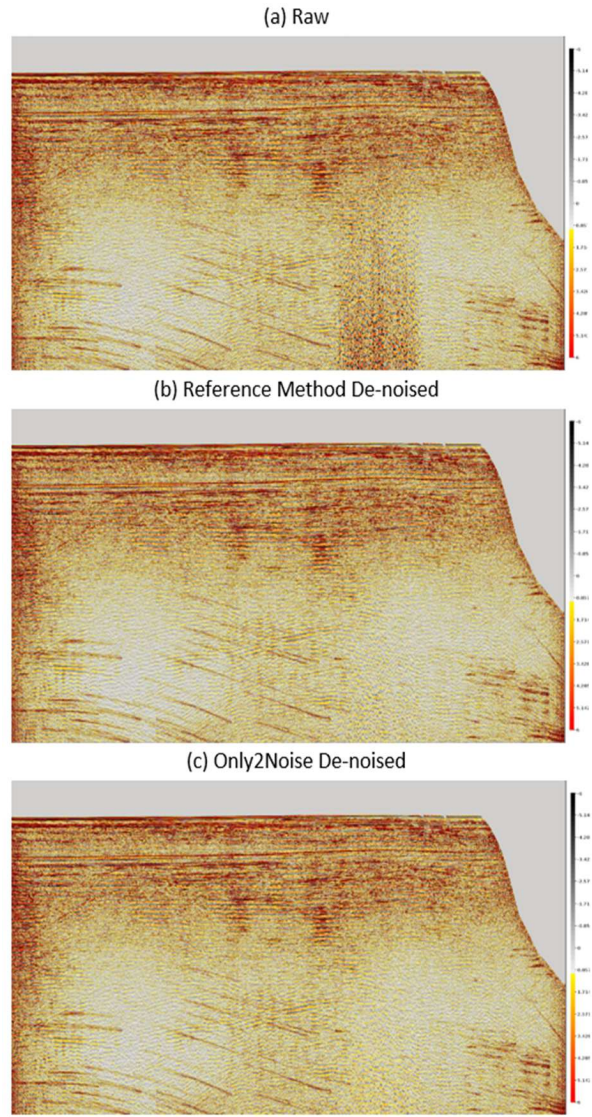

Figure 7. Field Data: Stack results comparison of corrupted raw data (b), Reference de-noised data(b), and Only2Noise de-noised data (c).

By detailed and careful checking of the frequency panels, common channels, and stacking results, our model indicates that the swell noise have been effectively detected and removed for a Gulf of Mexico asset. In respect of common channel (refer to Figure 6) and stacking results (refer to Figure 7), we could observe that Only2Noise method has strong ability of swell noise attenuation for seismic field data. Specifically, comparing with de-noised common channel results by reference method, referring to Figure 6 (b) and (c), Only2Noise model removes more swell noise in some parts (circled in red). Besides, comparing with the stacking results between de-noised results by the reference method and by Only2Noise method, we can see that our deep-learning-based method could reach a better ability for removing the swell noise. The results show that the proposed technique successfully returns an extremely premium imaging quality with very limited influence on the signal of interest.

The main contribution is to advice a deep-learning solution for the swell noise attenuation. Having rigorous comparisons with conventional methods for several benchmark studies, our proposed deep-learning models successfully implement the tasks with a great success, and achieve the prominent improvements in noise attenuation while the signal is preserved.

**Salt and Inclusion Integration with FWI for Seismic Imaging:**

Full-waveform inversion (FWI) has become a popular method to estimate elastic earth properties from seismic data, and it has great utility in seismic velocity model building and seismic reflectivity imaging in areas of complex salt. FWI is a non-linear data-fitting procedure that matches the predicted to observed waveform given an initial guess of the subsurface parameters. The velocity model parameters are updated to reduce the misfit between the observed and predicted data until the misfit is sufficiently small. Sharp velocity boundaries such as between salt and sediment are often updated manually for each iteration based on the seismic reflectivity images. Here, two different ideas and implementations of leveraging deep-learning approaches (Lu *et al*. 2018). to assist enhancement of seismic imaging quality through full waveform inversions are introduced below.

First, we propose the idea by designing a predictive neural network architecture as a potential alternative to mimic the complex FWI workflow, which is inspired from the unsupervised learning model of prediction of future frames in a video sequence. It is explored to directly simulate inversion procedures for seismic processing. Such neural network architectures are comprised of two main components: an encoder based on convolutional neural networks (CNNs), and a recurrent neural networks (RNNs) for iteratively predicting geophysical velocity models. Both the proposed networks could perform robust training for each individual layer and make a layer-specific prediction, which is compared with a target to produce an error term, which is then propagated to the subsequent network layers. With a several iterative training steps, the networks could learn internal representations decoded from latent parameters of seismic wave propagation that controls how FWI velocity modelling converges. These representations learned from one dataset could be transferred to predict the future velocity model of a brand-new area where the shape of salt body is not well imaged or known. Altogether, experimental results in figure 8 generated from a real Gulf of Mexico seismic data suggest that the inference, in representation of a powerful framework of unsupervised learning, which potentially provides an alternative approach to the FWI procedure, ultimately generates a sharp subsalt image with a high-resolution velocity model including an accurate salt model.

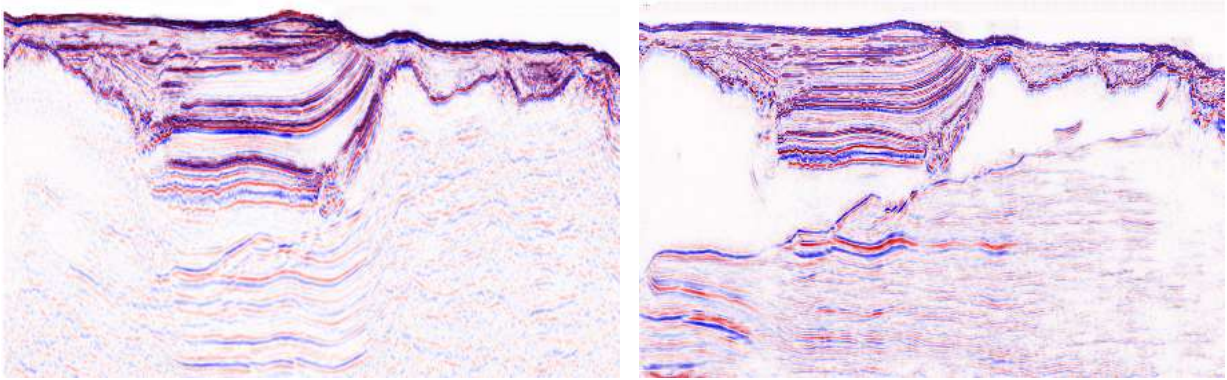

a. Interim inference-vague subsalt          b. final inference-clear subsalt

Figure 8. Deep-learning assisted FWI

Second, deep learning models are exploited to augment velocity model building workflows during 3D seismic processing in salt-prone environments. A new workflow integrated with deep learning model is introduced for enhancing the seismic imaging quality with improved process of generation of salt velocity models. There are two significant benefits of leveraging the proposed approach compared with conventional one. First, the salt and inclusion probability cube could be used as an extra regularization term integrated in the objective function. In order to find out the optimal direction for the optimization problem, searching in three dimensions is an extremely challenging and time-consuming task. In conjunction with output from deep learning mode, the searching area is perfectly regularized in the salt and inclusion probability mask region, which provides an opportunity of reaching the convergence with a faster pace. A 3D cube indicating possibilities of occurring salt bodies and inclusions, as shown in figure 9, is generated through ensembles of predictions from multiple models in order to reduce variances. Second, a salt velocity model, which could be used for initial or intermediate velocity model for FWI algorithm, can be directly converted from a salt and inclusion probability cube through inserting surrounding velocity model. A precise starting or intermediate velocity model is critical for optimization process, which may end up finding the local minimum instead of global minimum due to the initial guess far away from the ground truth. Through leveraging the velocity model inferred from deep-learning approach, figure 10 (b) shows the enhanced seismic imaging quality at the region beneath the salt-body where more continuous seismic reflections are observed.

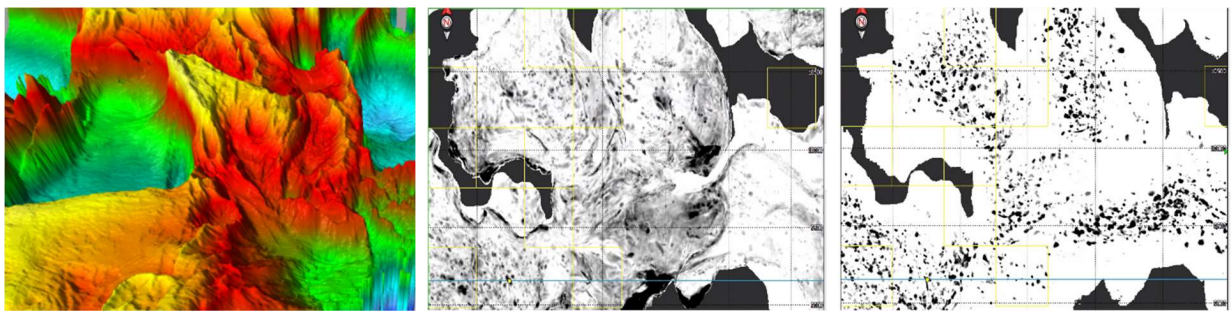

Figure 9. (a) Salt-picking     (b) Salt inclusions - conventional results   (c) Salt inclusions - DL results

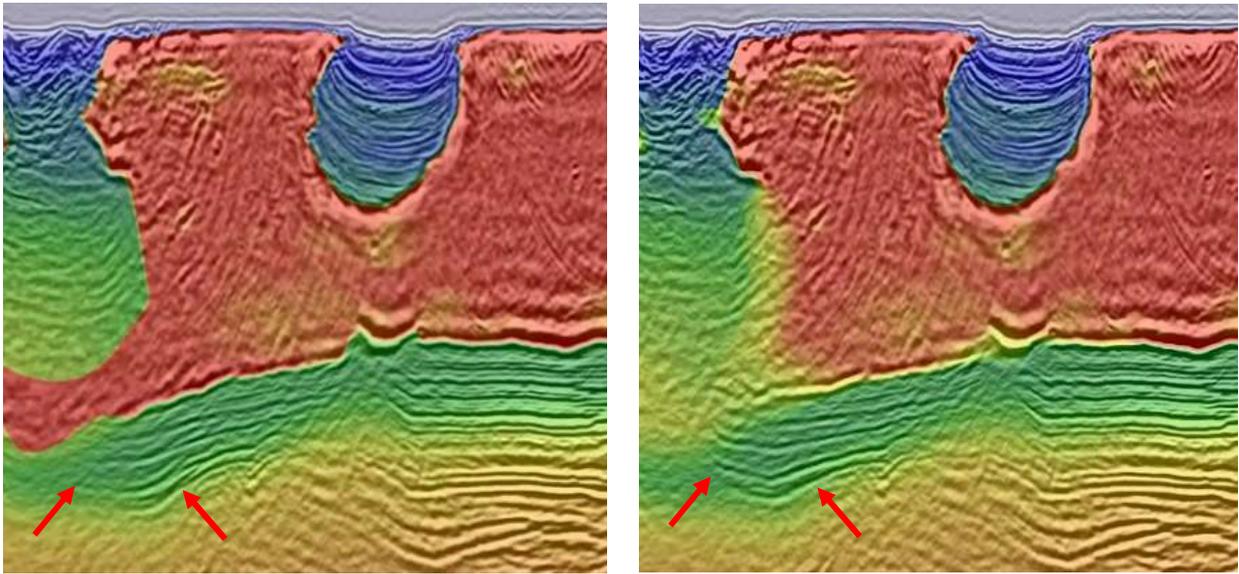

Figure 10. Enhanced seismic imaging for subsalt region with velocity model from conventional approach (a) and inferred from DL (b)

## Seismic Interpretation

**Fault Interpretation**

Fault-interpretation is considered as a significantly slow and tedious process with huge manual-efforts for large-scale seismic volumes. Deep-learning model has been successfully designed to assist geophysicists and geologists for this task.

One real use-case from the Gulf of Mexico offshore asset is selected to illustrate that the proposed approach could precisely capture and delineate 3D fault architectures in various spatial orientations. As shown in figure 11(a&c), the studied area is dominated by a series of horst and graben structures which resulted from multiple tectonic events and multiphase stress (extension, compression, translation), and contains complex faults across multiple zones at various depths. The developed novel model enables to handle such complex scenarios, and the interpretation with a 3D fault probability distribution cube brought by our algorithm shows vastly superior results in figure 11(b) with comparison to conventional interpretation which is usually induced from seismic coherence and semblance through a benchmarking comparison among several evaluations.

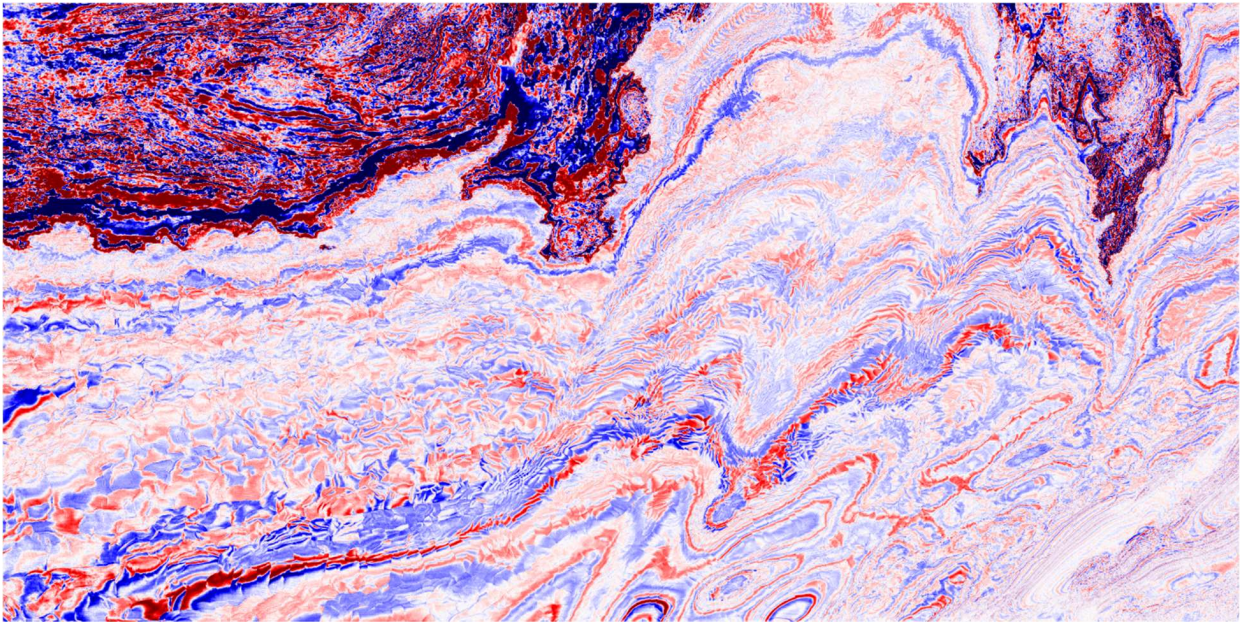

Figure 11(a). Depth View (raw seismic)

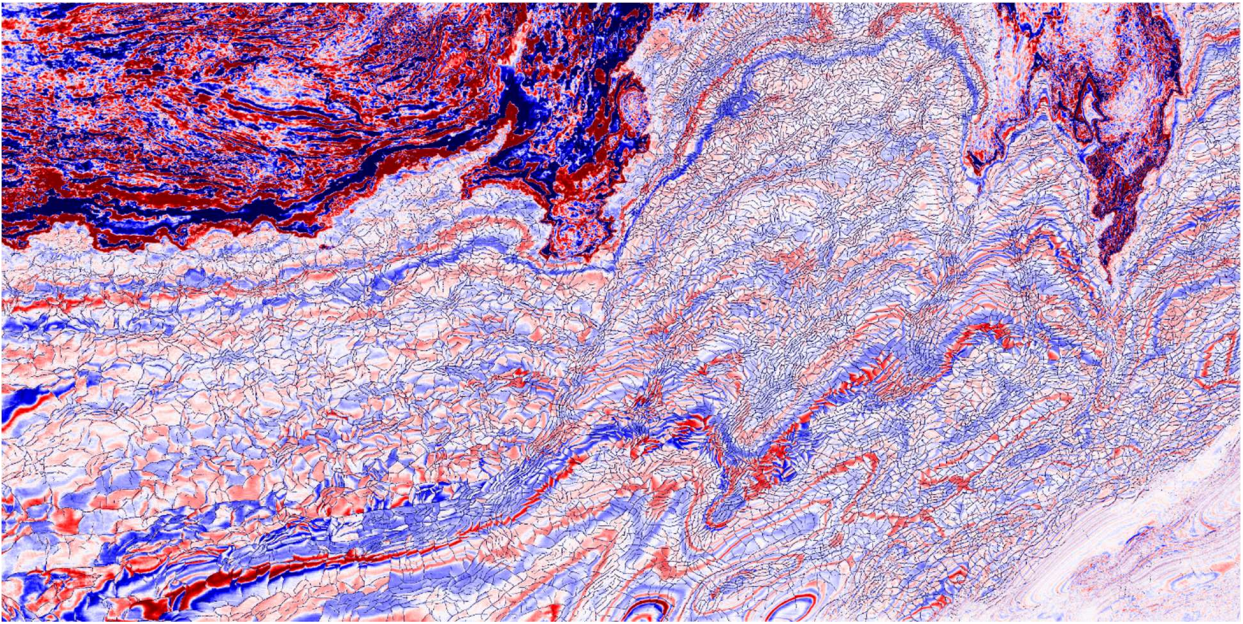

Figure 11(b). Depth View (faults inference)

Section view in figure 11(d) from the same area displays the robustness of the developed algorithm in detecting very large counts of deep, shallow and subtle faults that are extremely time intensive for Geophysicists to manually interpret.

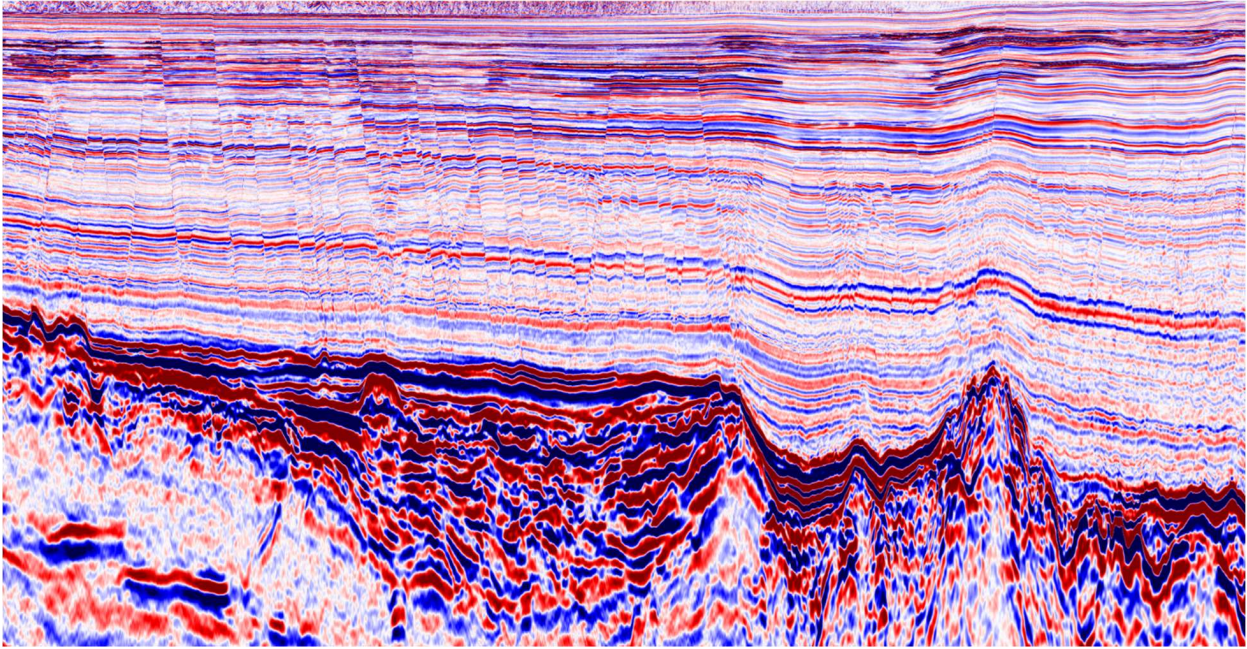

Figure 11 (c). Crossline View (raw seismic)

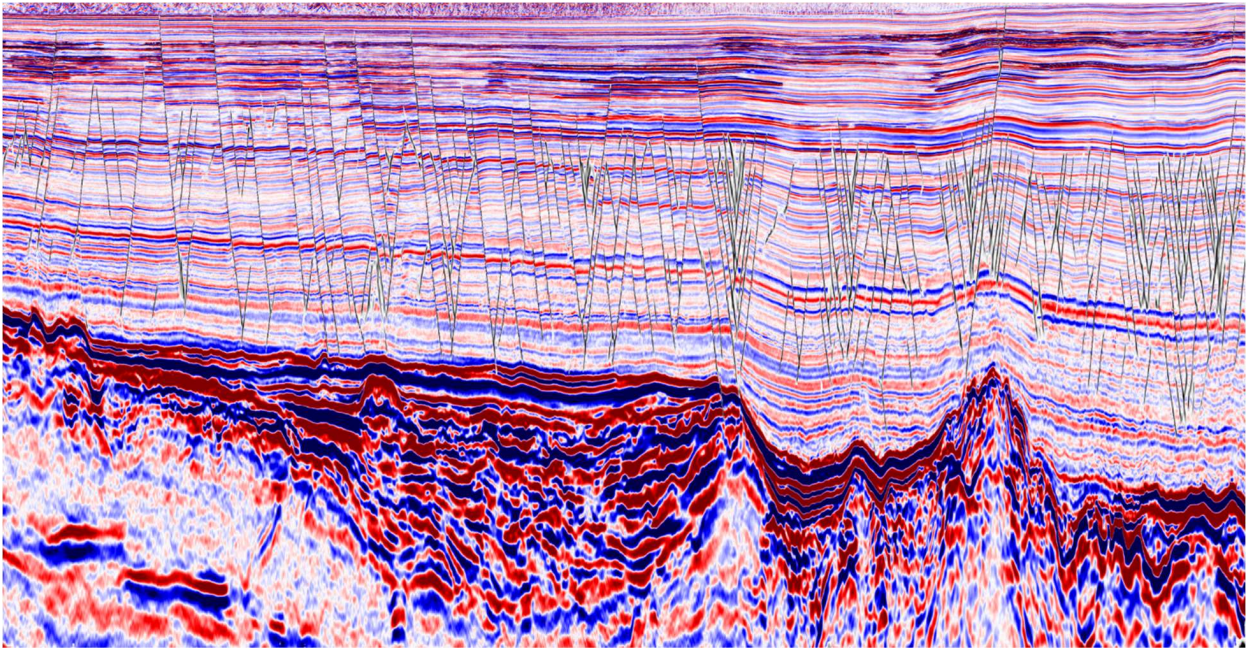

Figure 11 (d). Crossline View (faults inference)

We also demonstrate a refinement to a deep-learning fault identification process that improves the continuity and compactness of predicted fault planes in areas where faults intersect and quality of seismic images is relatively poor. Historically, predictions from both deep learning and traditional algorithmic approaches were characterized by "blurry" clouds of intermediate probability values that extended well beyond the fault plane. To remediate this blurring problem and enhance confidence of inferences, we demonstrate a preprocessing technique in the image domain by using GANs that sharpen the seismic image prior to training and prediction. This sharpening solution consists of two neural networks. A feature-extraction network is used for extracting both local and global features from an unrelated, high-quality "donor" seismic survey. Then, the data set of interest is sent through a donor reconstruction network where a generator architecture creates plausible-looking images at a denser sampling rate with high perceptual quality.

The example shown in figure 12 (Morris et al. 2019) presents a scenario from Gulf of Mexico region where deep, shallow and subtle faults have been detected in high fidelity regarding fault labelling continuity and fault tip terminations right beneath the base of salt-bodies where it is a challenging region for deep-learning model to precisely perform the task of detecting faults due to the relatively poor quality of seismic images with a low signal to noise ratio. For this experiment, we create our sharpening network using a modern, high-fidelity, deepwater 3D survey with well-imaged faults as the donor. The resulting generator architecture is then applied to our data set of interest — an altogether separate deepwater data set in the Gulf of Mexico. Similar in intent to a 5D interpolation, the GANs-based supersampled data contain three times the inline and crossline trace density, and the sampling interval is upsampled by a factor of three. This approach aims to preserve spatial and temporal frequency content of the parent data while providing a denser data set for deep-learning applications. The supersampled data is then deployed into our deep-learning training regimen to enhance the performance of our fault detection network. By introducing a preprocessing sharpening step, the predicted faults are less blurry, more compact, and more amenable to programmatic attempts to segment them into discrete features.

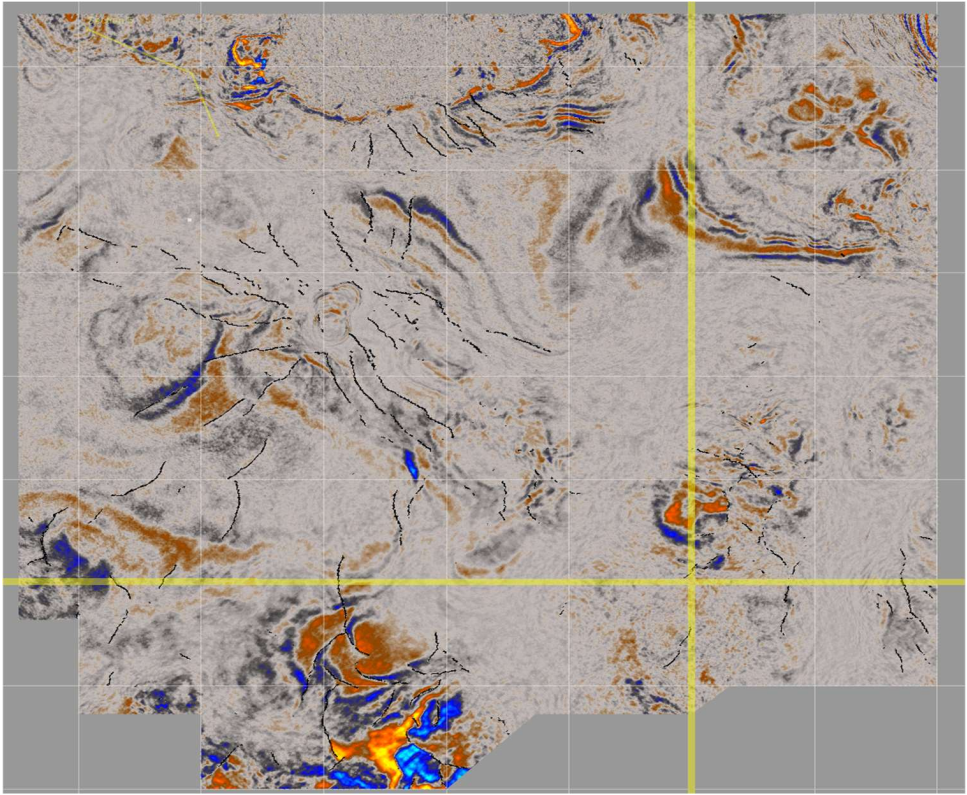

Figure 12 (a). Depth View

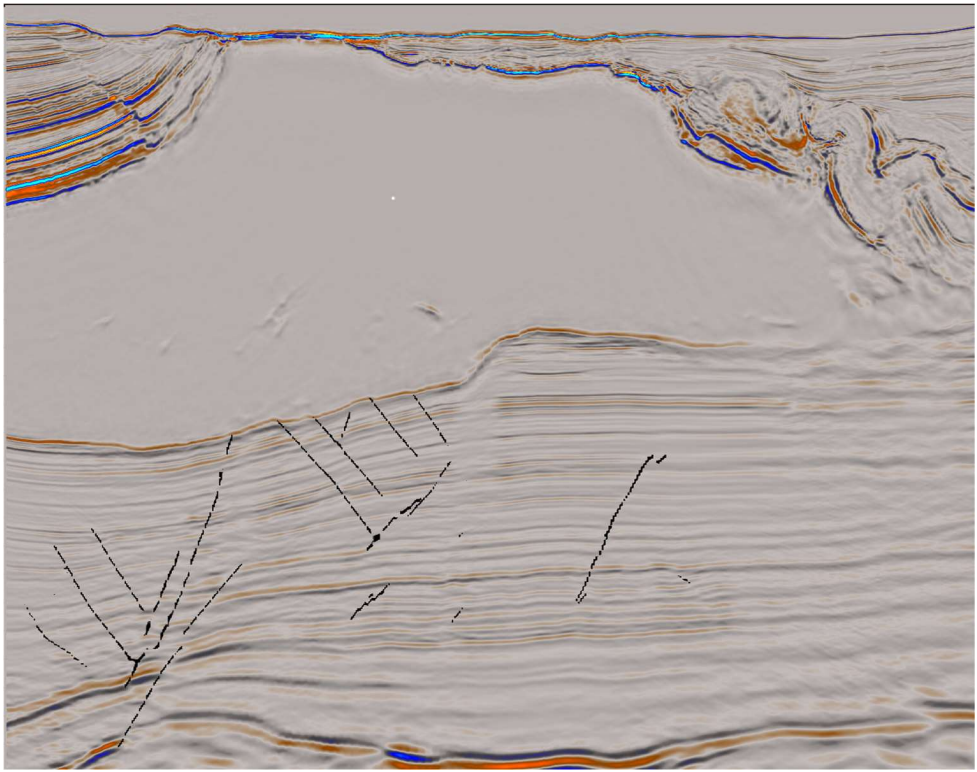

Figure 12 (b). Depth View

In summary, the developed algorithm has proven to add value to our exploration and production business in the oil and gas companies by significantly reducing the time required to evaluate the fault architecture of prospective geologic datasets, and has also greatly improved the fidelity of benchmarked fault interpretations from traditional methods.

**Channels Detecion by 3D Deep-Learning Model**

Deep water channels facilitate the transport of sediments into deep-water environments allowing sands to accumulate within slope and basin floor environments. These channel geobodies tend to accumulate into channel complex sets and are often considering primary targets for hydrocarbon exploration. Understanding the architecture and configuration of these channel complexes is a crucial step in defining the lateral and vertical continuity of many deepwater reservoirs. Individual channels may be prolific, erode previous deposition, aggrade vertically or laterally—often generating complex, intersecting geometries. Given the complexity and scale of these features, interpreters can spend weeks or months to delineate the channel complex morphologies. Their workflow traditionally relies on mapping sparse or discontinuous grids of seed picks that later are filled out using interpolation methods. Often, auto-tracking methods fail to reliably follow channel boundaries – as they are rarely represented by a continuous seismic loop.

We present a novel workflow to channel complex mapping that is based on deep-learning techniques, which isn't beholden to the limitations of conventional correlation-based tracking algorithms (Castillo *et al*. 2019). GANs and a deep discriminative network architecture are designed to detect channel complexes in the stratigraphic column. In addition, the technique allows multiple complementary attributes derived from raw seismic data to be consumed simultaneously during the training phase.

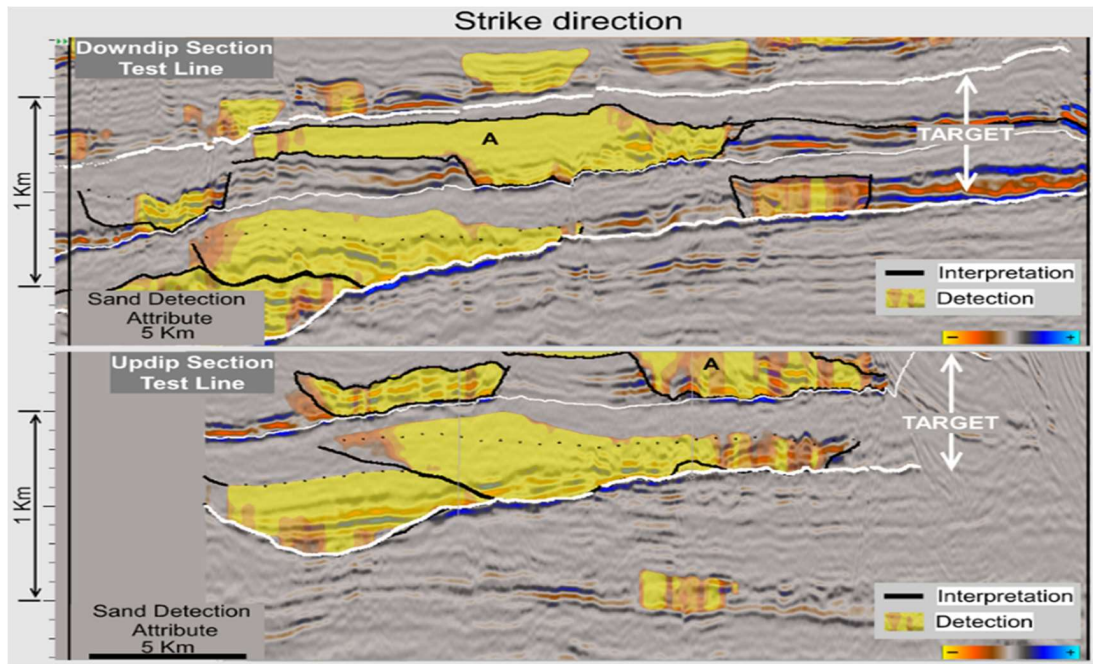

Figure 13. Detected channel complex probability VS handpicked interpretation.

A case study focuses on a deepwater 3D seismic survey. The imaged strata are dominated by multiple channel complex sets that punctuate zones of quiescent shales. We focus on an area of interest that extends over a depth window of 2000m and an area covering approximately 800km$^2$. Figure 13 shows the comparison between the detection probability volume (yellow/orange shaded areas) and the handpicked channel top and base surfaces (solid-black lines). The resulting trained model reliably handles inherent subjectivity and biases that are inherent to a stratigraphic interpretation. The associated top and base surfaces can rival an expert interpreter's manual efforts. Since it is a data-driven operation at each trace, it obviates the need for spatial interpolation techniques to 'fill the gaps'. The technique deepens the set of interpretation tools available to track surfaces across a seismic dataset. These machine-derived 3D geobodies can then be post-processed to extract traditional top and base surfaces, if so desired. Deep learning-based workflows offer the potential to deliver more precision and speed to the interpreter who sets out to map deepwater channels, without sacrificing the accuracy associated with traditional grid-based mapping methods.

**Horizon Picking**

Horizon picking in 3D seismic data is a very challenging task, however, it plays an important role in understanding the geologic information. The popular techniques of automatic extraction for horizons have several limitations: (1) the selection of picks in a trace usually ignores lateral continuity, and (2) the trace traversal order can result in significantly different horizons in a consequence that the resulting picks in the same horizon often conflict with each other. Thus, we introduce a deep-learning approach to detect horizon surfaces from 3D seismic data. The example shown in figure 14(a) covers a large area involved with complex geologic varieties, and it elaborates that the deep-learning approach enables to generate several horizon surfaces not only in an efficient way with providing very few training labels, but also in an accurate way with continuous 3D horizon surfaces in separating relative geologic time for interpretation purpose, as shown in figure 14(b). By incorporating deep-learning models, the time to evaluate horizons for entire dataset has been significantly reduced from months to days to better assist our asset team in the decision-making process.

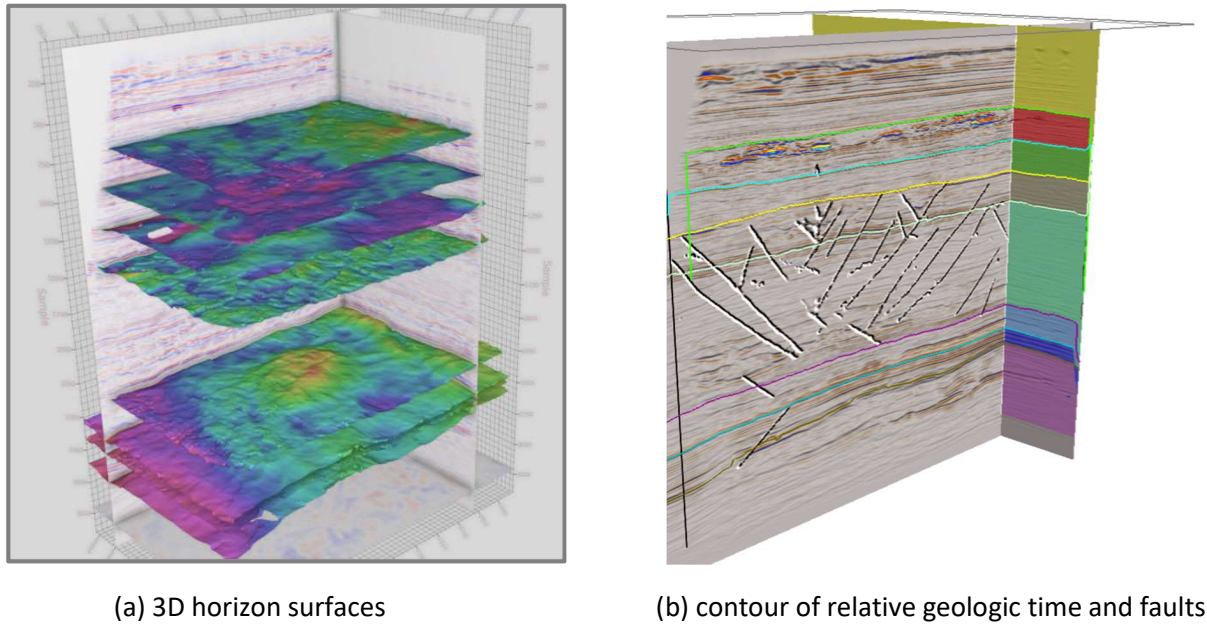

(a) 3D horizon surfaces  (b) contour of relative geologic time and faults

Figure 14. Horizon picking from Volve Dataset

## Seismic Inversion

**Sand thickness detection:**

Due to the inherit nature of seismic data, precise identification of the reservoir characteristics is an extremely challenging task. We successfully design a deep-learning model integrated with synthetic wedge models to overcome the geophysical limitations while performing the interpretation of thickness of sand bodies from a low resolution of seismic data (Lu *et al*. 2019). Through understanding and learning the geophysical relationship between seismic responses and corresponding indicators for true sand thickness, the deep-learning model could automatically detect the locations of top and base of sand bodies identified from seismic traces by precisely revealing the lithology distribution. Quantitative analysis and extensive validations from wedge models and field data prove the robustness of the proposed methodologies. The true sand thickness, identified from the deep-learning model, provides an extremely useful guidance in enhancing the interpretation of lithological and stratigraphic information from seismic data. In addition, the proposed deep-learning approach eliminates the risks of over- and under-estimation of net-to-gross with a significant improvement with respect to the accuracy.

Figure 15 shows one example with a region where the deep-learning model adjusts the seismic thickness to both scenarios by thickening and thinning of sand thickness identified from seismic trace, which demonstrates the solid evidences that the deep-learning model has learnt from training with the wedge model by gaining geophysical insights from seismic amplitudes regarding corrections of below and above tuning thickness.

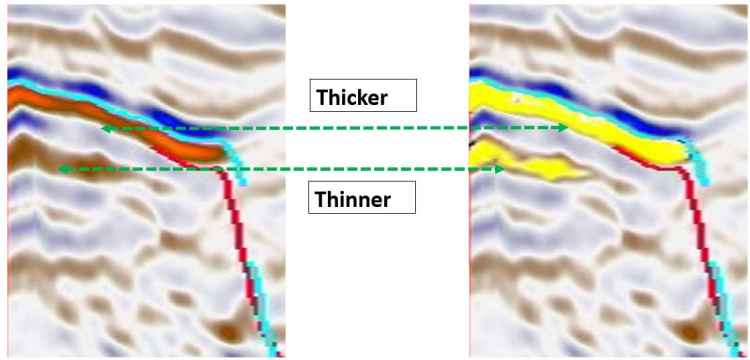

Figure 15. Inference on Field Data

Even with significant interferences from seismic traces, the proposed approach achieves superior interpretability with a strong link from seismic amplitude to an indicator of sand thickness, which introduces a much easier and more accurate approach in identifying reservoir thickness and characterizations. The isochrones, generated from deep-learning outputs, provide finer details and higher resolutions, being able to draw a clear portrait for sediment deposition of true thickness as shown in Figure 16.  It proves that the proposed deep-learning approach could more effectively handle a large 3-D seismic volume with consistent criteria as well as to infer thickness by avoiding tuning effects in a totally automatic manner with much higher accuracy than human. With fewer interpretation artifacts and interference fingerprints, seismic-guided stratigraphic profiling and depositional-facies could be successfully revealed, when deep-learning technique in conjunction with wedge model are applied.

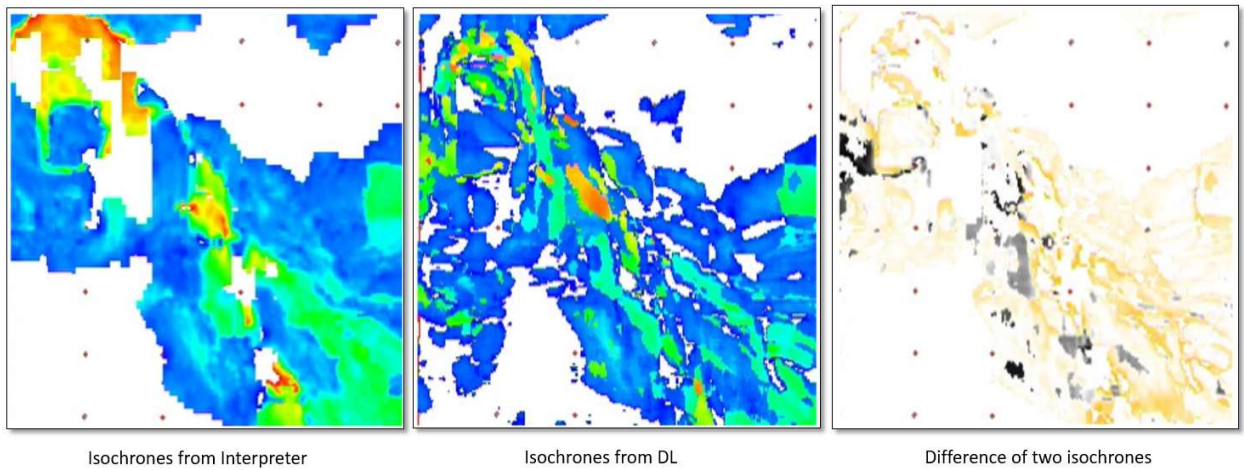

Figure 16.  a) Isochrones generated from an interpreter based on seismic information. b) Isochrones generated from deep-learning. c) Difference between two isochrones.

**Enhancement of Seismic Imaging**

Enhancing the frequency bandwidth of the seismic data is always the pursuance at the geophysical community. High resolution of seismic data provides the key resource to extract detailed stratigraphic knowledge. A novel approach, based on deep learning model, is introduced by extracting reflection amplitudes from Elastic Impedance (EI), which contains extremely rich high and low frequency information, to broaden spectrum bandwidth of seismic data through boosting low and high components (Zhang *et al*. 2019). The corresponding improvement is observed from the new attribute, which is inferred from the deep-learning model and is named as synthetic EI, with the enhanced resolutions from the seismic data, due to boosted low and high frequency parts, as well as elimination of sidelobe artifacts from the seismic amplitudes. During the training stage of deep learning model, geo-spatial information by taking consideration of multiple wells simultaneously is fully guaranteed, which assures that laterally and vertically geological information are constrained by and accurate away from the well controls during the inversion procedure. Extensive experiments prove that the enhanced attribute is consistent with well log information, and honors rock property relationships defined from the wells at given locations. Uncertainty analysis could also be quantitatively assessed in order to determine the possibilities of a range of seismic responses by leveraging the multiple realizations from the proposed approach.

Figure 17(a)(b) illustrates the wavelet and spectrum comparison between the original seismic data (blue) and synthetic EI inferred from deep-learning model (green). Figure 17(a) presents that the peak of the original seismic amplitude is compressed to be more centralized from the observation of the synthetic EI as well as the phenomenon of sidelobe effect has been eliminated quite significantly comparing to the original seismic data. In addition, the spectrum analysis in Figure (b) of the original seismic and synthetic EI shows that the deep neural network not only boosts the high frequency components of the seismic, but also generates more low frequency information to the synthetic attribute.

Figure 17(c) shows a comparison between real logs versus generated broad bandwidth trace. The synthetic EI matches the EI quite well, such that both detailed spikes and curve are captured, which proves the claim that the proposed approach is capable of not only extending a broader bandwidth from seismic volume, but also capturing different rock properties.

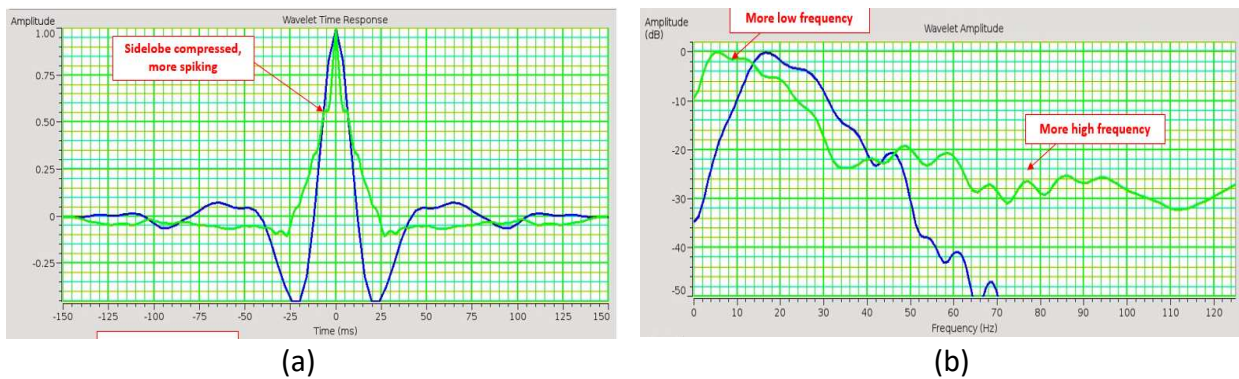

(a)          (b)

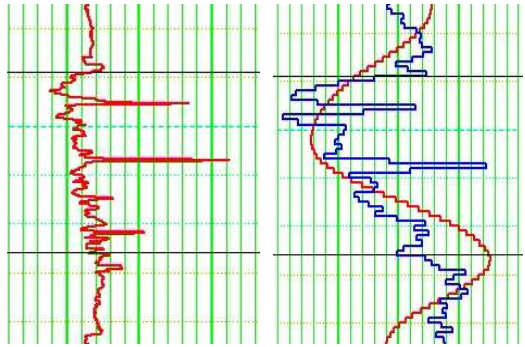

(c)

Figure 17. (a) Comparison between seismic trace and its corresponding synthetic broad frequency signal outputted from DL network (b) Spectrum analysis (c) Comparison between real logs versus generated broad bandwidth traces

Figure 18 displays the difference between original seismic and the generated outputs from our proposed deep-learning model in depth slice. Our broad bandwidth seismic maintains the geology features from input information, such as the faults in original image remain unchanged. In the meanwhile, the generated seismic image with broad spectrum has high resolution, revealing more detailed deposition information and clear boundaries between various facies.

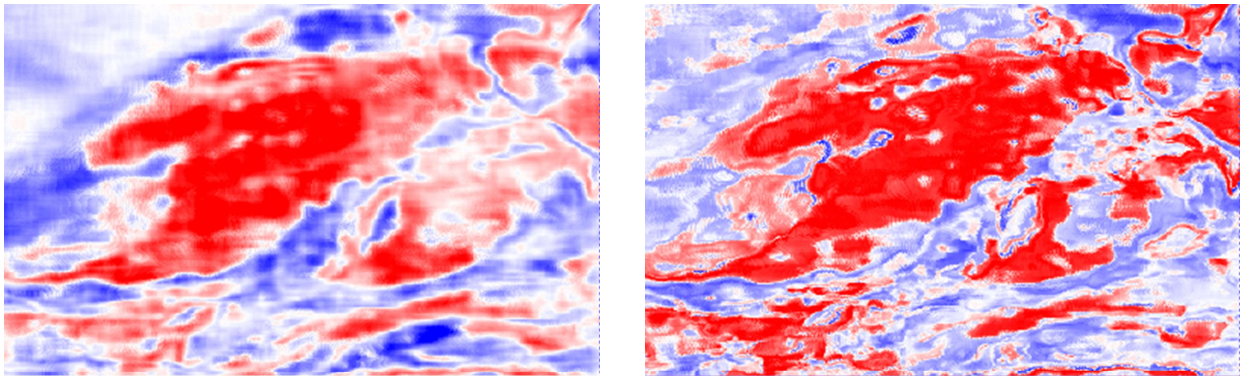

Figure 18. Original seismic versus broadband seismic from DL network in depth slice

**Multi-facies inversion:**

A deep-learning based solution is introduced with the following three steps to reveal multi-facies properties on a fine scale from the inversions of extensive seismic surveys. First, GANs, introduced at the previous session, is leveraged to recover high and lower frequency signals, extracted from elastic impedance, back to seismic traces for the entire survey. Second, a deep encoder-decoder is designed to classify various rock facies along each single trace with low and high frequency recovered from previous step. As a result, by combining GANs with encoder-decoder, deep-learning models could successfully provide probabilistic rock types for each location in the seismic volume. Third, the inversion inevitably yields uncertainties due to the nature of the generative model and the non-uniqueness of seismic inversion. These uncertainties can be easily manifested as unreliable inference of the subsurface properties. So, the question is how to validate the prior information of high-resolution inversion solutions generated from the previous step. We are proposing an end-to-end solution to close the loop. In our workflow, the rock physics modelling is used to calculate elastic rock physics properties for the potential rock types inferred by deep-learning models. They are then used to generate a set of synthetic seismic traces to be compared with the real seismic traces. Eventually, all matching synthetics reveal validated rock type candidates. In such a way, results from deep-learning models are perfectly validated by the real seismic observations. The loop is closed by starting from seismic and then ending with seismic in an efficient manner by skipping generations of unnecessary random guesses from the conventional inversion approach, like Markov-chain Monte Carlo (MCMC) algorithm. With this three-steps workflow, we are able to achieve facies distributions in the target region.

The novel hybrid seismic waveform inversion framework, incorporating generative and discriminative models to discover earth properties in fine scale with geophysical modeling to filter out artifacts, is demonstrated to achieve state of the art for inversion science, with a real-world case of seismic inversion from Africa to present the robustness and effectiveness of the proposed approach. Figure 19 (a) presents a binary classification of either sand or shale where red color indicates a higher probability of being rock versus blue color indicating a higher chance of being shale. Figure 19 (b) elaborates three different rock properties, represented with dark, light blue, and green colors, and two shale properties, represented with orange and red colors, have been determined by the proposed deep-learning approach.

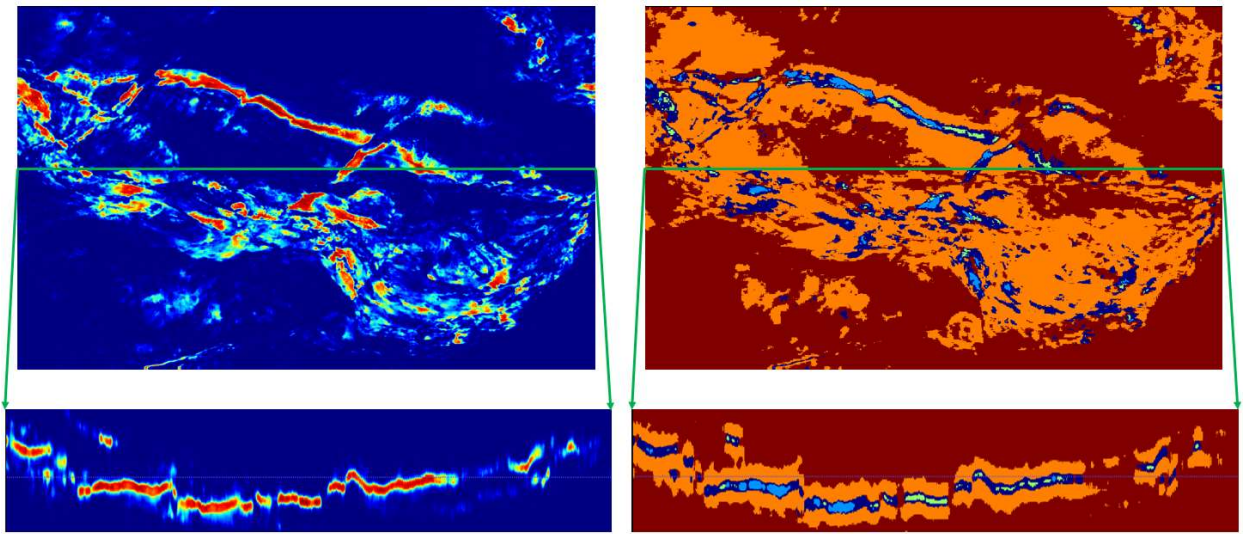

Figure 19. Facies classification

**Deep Neural Network Architectures**

Deep neural networks could play an important role in assisting tasks for seismic acquisition, processing, interpretations, and inversions, etc, which could release significant amount of time from geophysicists in manually doing such work. However, it is not an easy task to build effective deep-learning models for these tasks mentioned above.

**Design of Discriminative Model**

A deep discriminative model with convolutional encoder-decoder architecture, equipped with a powerful classifier, is considered for interpretation tasks, where the encoder portion extracts feature maps from the images, and the decoder attempts to semantically project the discriminative features (lower resolution) learnt by the encoder onto the pixel space (higher resolution) to return a pixel-wise classification. Stacks of convolutional filter kernels are used to benefit the task of segmentation. Dilated convolution, which increases the receptive field without downgrading resolution, is considered during the network architecture design. Nonlinear activation functions (ReLU) are used as a mapping function to transform a multidimensional input image into a desired output. Batch normalization, which reduces the change in distribution of network activations, is also used between the convolutional and ReLU layers improve convergence during training. The ReLU activation layers are followed by a maxpooling operation, effectively reducing each channel of activations. The number of activation channels in each subsequent layer is progressively increased to capture increasing feature complexity. Random crops are performed to help in reducing the effects of class imbalance. Only very few training labels, consisting of about 1 to 2 percent of total area, are required and prepared by geophysicists interpreting the 3D seismic data. Enhanced image augmentation techniques have been deployed in order to generate additional training data. Special care for optimum selections of optimization methods, learning parameters, and model hyperparameters is needed specially for seismic images. Transfer-learning technique is considered during training process for a new geologic area which was not exposed previously to the neural networks.

Since most of the geo-bodies, which need to be indented from seismic data, are in 3-dimentional format. Therefore, a 3D deep-learning model, which is able to take care of lateral and vertical information, is preferred. Specifically, a neural network architecture, with 3D convolutional, de-convolutional layers, and 3D max-pooling, is designed to take standard amplitude of 3D seismic volumes as an input. Enhanced data augmentations through generative adversarial networks and a weighted loss function enable the network to train with few sparsely annotated slices. In addition, multiple attributes, which could provide complementary information to support geologic features, can be simultaneously adopted and consumed by neural networks during the training process.

The spatial consistency in salt body detection is barely addressed from 2D approach in general, while this is a remarkable aspect of the 3D approach. The comparison shown in figure 20 presents the 3D approach explicitly provides spatially consistent results with continous geo-bodies.

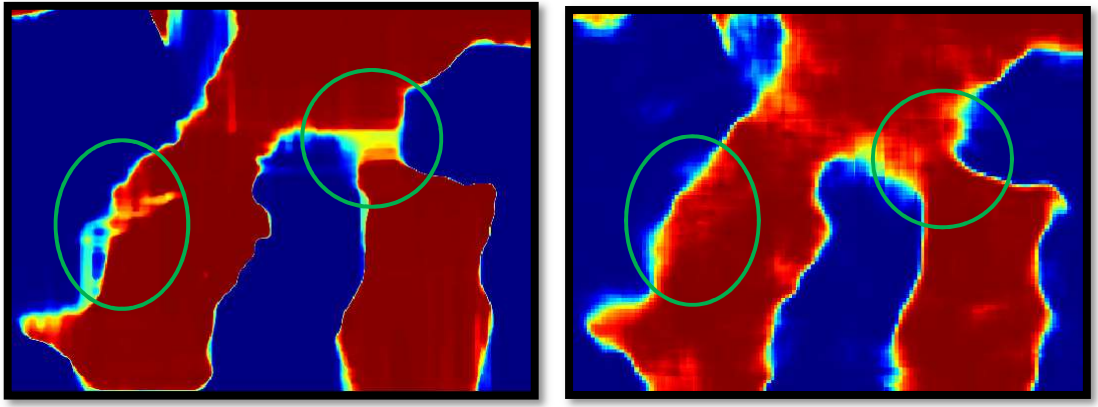

Figure 20. Comparison of geo-bodies in depth between 2D and 3D models

**Enhancement of Performance of Seismic Interpretations**

Generative Adversarial Networks (GANs) has been considered as the most innovative breakthrough in the generative model domain, which has been extremely popular and successful in the computer vision domain. A successful application of GANs to enhance the sampling density of real seismic data is presented (Lu *et al*. 2018). This approach provides greater perceptual quality of the native seismic data, preserves the spatial and temporal frequency content, and introduces no distortion of amplitude extrema potentially impacting quantitative evaluations. This additional step is applied and tested on interpretation of faults to improve the outcome from an existing deep-learning-based detection network. The resulting faults are eventually detected with greater clarity and precision. The proposed methodologies, by leveraging GANs in combination with a discriminative model equipped with a powerful classifier, enables precise machine-derived mapping of complex fault networks where faults intersect or trend in close proximity to one another as shown in figure 21.

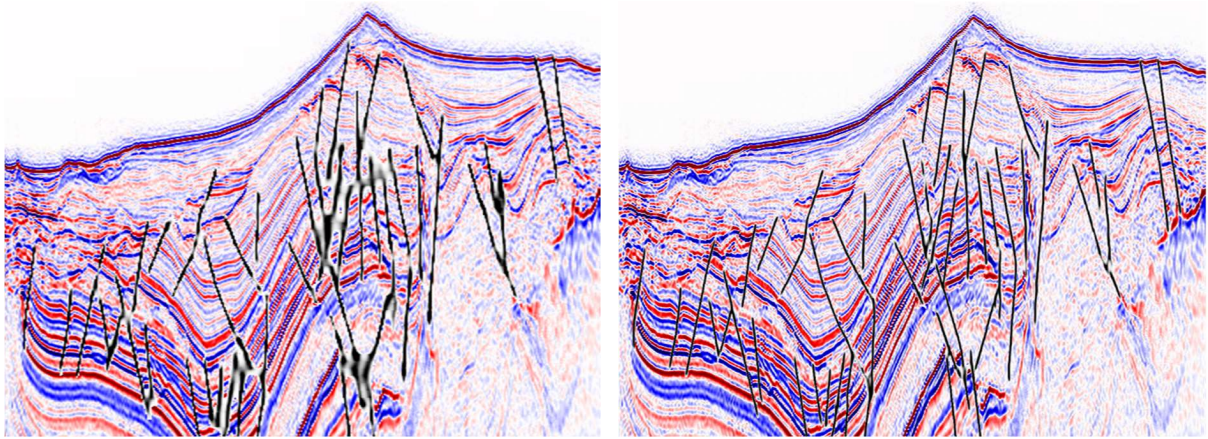

Figure 21. Comparison between (a) native data and (b) supersampled data used for a deep-learning approach to fault detection.

# Generalization of Deep-Learning Model

Development of a global deep-learning model, training with previously existing labels within one region and being able to successfully perform interferences on another independence survey, is highly desirable. In other word, the eventual goal is to avoid providing any localized labels to the deep-learning model, but it still enables to deliver results in similar quality to the ones using the standard pipeline training with local geological contexts. The ability to make an inference on a totally new region, where the deep-learning model hasn't been trained before, requires a strong capability in terms of generalization of the deep-learning model.

However, generalization of deep-learning model is still an open problem in the computer vision domain. Given the fact that different seismic datasets may equip with various frequencies, noise levels, migration and inversion algorithms, scale ratios, etc., regularization of seismic data is a vital step for achieving the goal. Here, GANs is proposed to facilitate the regularization of seismic data across surveys such that deep discriminative models, equipped with a powerful classifier, could train and make the inferences on uniform seismic datasets, regardless of complexities of geologies and differences from acquisition and processing stages.

Another limitation that constraints the capability of generalization of deep learning model is the lack of training labels, which is an extremely time-consuming task for geophysicists to prepare. Thus, the preparation of training dataset representing as much variations of geological information as possible will be helpful in reaching a global deep-learning model being able to make an effective inference. Here, GANs are leveraged to generate plausibly synthesized seismic images and corresponding labels. Through learning density distribution of seismic datasets representing from different regions, GANs can be used to extract geological features, and combine prior knowledge to generate plausibly synthesized seismic data. Figure 22 shows four synthetic seismic images, which are hard to distinguish from real seismic images, including rich geologies all generated by GANs. By providing a chance to cover varieties of geologies through synthetic labels and paired synthetic seismic images, the deep discriminative model is able to successfully achieve the generalization purpose. In addition, it could also benefit to reduce complexity of networks in terms of number of parameters, enhance the stability of optimization algorithm, and improve robustness of adversarial perturbations.

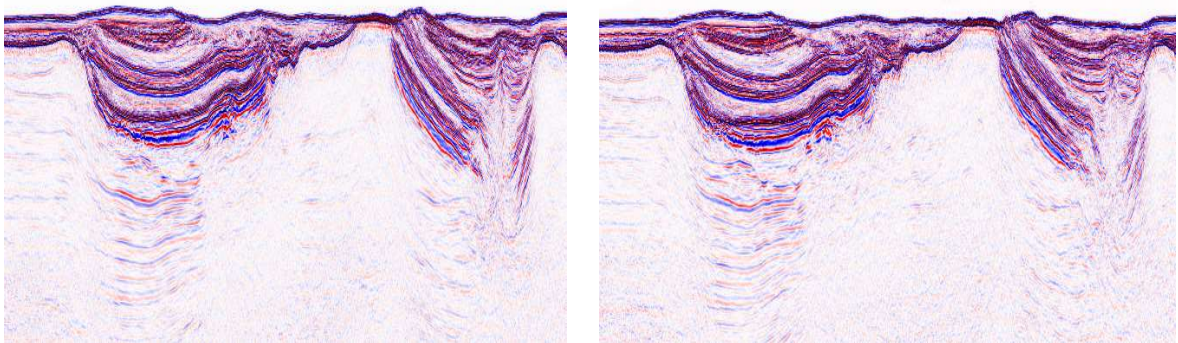

(a)             (b)

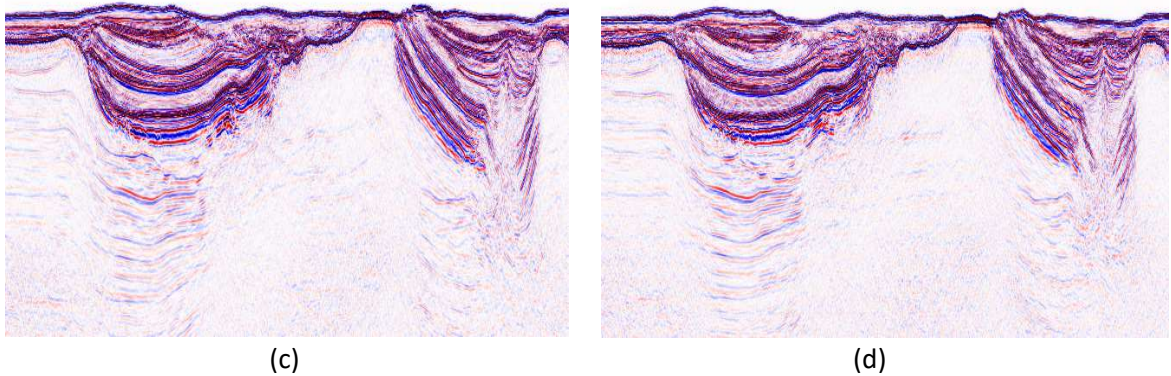

(c)　　　　　　　　　　　　　　　　(d)

Figure 22. Synthetic seismic images

A successful story, from Gulf of Mexico offshore asset selected as evidences to support aforementioned points, has launched by directly applying the weights learnt from one survey to successfully infer the faults on another survey with totally new geologies. The results look promising with directly skipping the training process.

Therefore, the combination of GANs, used for regularization and synthesis of seismic image and pair labels, and the discriminative model, enables a successful tool for automation of seismic interpretations. By having a consistent regularization across different seismic datasets and extra geological information, the deep-learning based technology, without the needs of further training, could easily reach the target of severing for large scale of users in the geophysical community who are trained with limited data science knowledges.

## Summary

The successful applications of deep-learning model for the geophysical topics shown in this manuscript builds up strong confidences and values by integrating deep-learning approach into classical tasks relevant to seismic interpretations and imaging. There are still many subjects, which are not mentioned in this manuscript, waiting to be solved from researchers in computer visions and geophysics to collaboratively work together by designing appropriate neural network architectures. In addition, meta-learning, active learning, and reinforcement learning are considered as popular branches within the machine-learning community, which are worth the efforts in developing and applying them to the geophysical domain.

## Acknowledgements

The author would like to thank Anadarko for permission to publish this work. In addition, special thanks go to my colleagues, Seth Brazell, Karilys Castillo, Jianxiong Chen, Cody Comiskey, Hunter Danque, Jeremy Graybill, Jon Koeing, Rebecca Li, Nikolaos Mitsakos, Matt Morris, Stan Morris, Christian Noll, Sanjay Paranji, Mike Seeber, David Sixta, Pennie Taylor, Yuan Xiao, Grace Yu, Yanyan Zhang, Xing Zhao, whom I closely collaborated with and I greatly received support from, during the Research & Development activities. Lastly, I would like to thank CGG, TGS, and WesternGeco for allowing the reproduction of their data.